\documentclass[letter,traditabstract]{aa} 
\usepackage{graphicx}
\usepackage{mydef}
\usepackage{multirow}
\usepackage{txfonts}
\begin{document}
   \title{High-resolution absorption spectroscopy of the OH $^2\Pi_{3/2}$ ground state line}
   \author{H. Wiesemeyer\inst{1}
          \and
          R. G\"usten\inst{1}
          \and
          S. Heyminck\inst{1}
          \and
          K. Jacobs\inst{2}
          \and 
          K.M. Menten\inst{1}
          \and
          D.A. Neufeld\inst{3}
          \and
          M.A. Requena-Torres\inst{1}
          \and
          J. Stutzki\inst{2}
          }
   \institute{Max-Planck-Institut f{\"u}r Radioastronomie,
              Auf dem H{\"u}gel 69, 53121 Bonn, Germany \\
              \email{hwiese@mpifr.de}
         \and
              I. Physikalisches Institut, Universit{\"a}t zu K{\"o}ln,
              Z{\"u}lpicher Str. 77, 50937 K{\"o}ln, Germany
         \and
              The Johns Hopkins University, 3400 North Charles St. Baltimore,
              MD 21218, USA
             }
   \date{Received ; accepted }
  \abstract
  {
   The chemical composition of the interstellar medium is determined by gas
   phase chemistry, assisted by grain surface reactions, and by shock chemistry.
   The aim of this study is to measure the abundance of the hydroxyl radical
   (OH) in diffuse spiral arm clouds as a contribution to our understanding of
   the underlying network of chemical reactions.
   Owing to their high critical density, the ground states of light hydrides
   provide a tool to directly estimate column densities by means of absorption
   spectroscopy against bright background sources. We observed onboard the SOFIA    observatory the $^2\Pi_{3/2}, J=5/2 \leftarrow 3/2$ 2.5~THz line of
   ground-state OH in the diffuse clouds of the Carina-Sagittarius spiral arm.
   OH column densities in the spiral arm clouds along the sightlines to W49N,
   W51 and G34.26+0.15 were found to be of the order of
   $~10^{14}\,\mathrm{cm}^{-2}$, which corresponds to a fractional abundance of
   $10^{-7}$ to $10^{-8}$, which is comparable to that of H$_2$O. The
   absorption spectra of both species have similar velocity components, and the 
   ratio of the derived H$_2$O to OH column densities ranges from 0.3 to 1.0.
   In W49N we also detected the corresponding line of $^{18}$OH.}
   \maketitle
%
\section{Introduction}
The first molecule discovered in space was the CH radical (Dunham 1937). At
that time, the first predictions were made regarding the abundances of
molecules in interstellar space, identifying the hydroxyl radical (OH) as a
promising candidate (Swings \& Rosenfeld 1937). It took more than two
decades until the latter was detected in the {\sc ism} (Weinreb et al.
1963), in absorption against the supernova remnant Cas A, in the $F= 1 - 1$
and $2 - 2$ transitions between the hyper-fine structure split
$\Lambda$-double levels of OH's  $^2\Pi_{3/2}, J = 3/2$ rotational ground
state. Comprehensive models of the gas phase chemistry of diffuse 
interstellar clouds constructed by van Dishoeck \& Black (1986) revealed
the importance of the OH radical in the network of reactions leading to the 
formation of oxygen-bearing molecules.
Unfortunately, OH column densities in these objects are difficult to
determine from radio lines, due to frequently observed deviations of the
underlying level population from LTE (e.g., Neufeld et al. 2002). The highest
densities occurring in diffuse clouds amount to $\sim$10$^4\,{\rm cm}^{-3}$
(Greaves \& Williams 1994), while their mean density is
$\sim$10$^2\,{\rm cm}^{-3}$ (Snow \& McCall 2006; Cox et al. 1988 derived upper
limits of a few thousand ${\rm cm}^{-3}$).

Storey et al. (1981) first detected the $\Lambda$ doublet line from the 
OH ground state ($^2\Pi_{3/2}\,J=5/2 \leftarrow 3/2$) towards Sgr~B2 near the
Galactic centre with the Kuiper Airborne Observatory, but their spectral
resolution (250~\kms) proved inadequate to separate the absorption by the
line-of-sight clouds from that occurring in Sgr~B2 itself. Here we report the
detection of one doublet line of this transition (at 2514~GHz) and of its
isotopolog $^{18}$OH (at 2495 GHz) with GREAT\footnote{GREAT is a
development by the MPI f{\"u}r Radioastronomie and the KOSMA/Universit{\"a}t
zu K{\"o}ln, in cooperation with the MPI f{\"u}r Sonnensystemforschung and the
DLR Institut f{\"u}r Planetenforschung.} onboard SOFIA, in absorption towards
the giant H{\sc ii} regions W49N and W51 and the ultracompact H{\sc ii} region
G34.26+0.15. Both lines are inaccessible for Herschel/HIFI. The three observed
lines of sight are within a $15^\circ$ wide Galactic longitude interval. Those
towards W51 and G34.26+0.15 cross the near side of the Carina-Sagittarius
spiral arm, with W51 ($\ell = 49\fdg5$) at a distance of
5.41~($+0.31,\,-0.28)$ kpc (Sato et al. 2010) and G34.26+0.15 at $\sim$2~kpc
distance (cf. measurements of G35.10-0.74, which has a comparable radial
velocity, Zhang et al. 2009).
The line of sight to W49N ($\ell = 43\fdg 2$) first crosses the near side of    the Carina-Sagittarius arm, grazes the Crux-Scutum arm, and then again crosses
the Carina-Sagittarius arm on its far side, where W49N is located at a 
distance of $(11.4 \pm 1.2)$\,kpc (Gwinn et al. 1992). 
%
\section{Observations, data reduction and analysis}
The observations reported here were performed with the GREAT receiver
(Heyminck et al. 2012) onboard the SOFIA airborne observatory (Young et 
al. 2012), as part of the {\it basic science} programme (flights on 2011 July 26
and November 8). On the first flight the receiver's M and L2 bands were tuned to
2514.317~GHz for the $^2\Pi_{3/2}\,J=5/2\leftarrow 3/2$ group of OH hyperfine
structure lines (in the lower sideband) and to 1837.817 GHz for the
$^2\Pi_{1/2}\,J=3/2 \leftarrow 1/2$ lines (in the upper sideband), respectively.
On the second flight, the M band was also tuned to the 2494.695~GHz frequency
of the $^2\Pi_{3/2}\,J=5/2\leftarrow 3/2$ transition of $^{18}$OH.
   \begin{table}
      \caption[]{List of the observed $^2\Pi_{3/2} J=5/2 \leftarrow 3/2$ $^{16}$OH and $^{18}$OH lines}
         \begin{tabular}{lll}
            \hline\hline
            \noalign{\smallskip}
            Transition   &  Frequency [GHz]$^{\rm (a)}$ & A$_{\rm E}$ [s$^{-1}$]$^{\rm (b)}$ \\
            \noalign{\smallskip}
            \hline
            \noalign{\smallskip}
            \multicolumn{3}{c}{OH, $^2\Pi_{3/2}, J=5/2 \leftarrow 3/2$} \\
            \noalign{\smallskip}
            \hline
            \noalign{\smallskip}
            $F=2^- \leftarrow 2^+$ & 2514.298092 & 0.0137  \\ 
            $F=3^- \leftarrow 2^+$ & 2514.316386 & 0.1368  \\
            $F=2^- \leftarrow 1^+$ & 2514.353165 & 0.1231  \\
            \noalign{\smallskip}
            \hline
            \noalign{\smallskip}
            \multicolumn{3}{c}{$^{18}$OH, $^2\Pi_{3/2}, J=5/2 \leftarrow 3/2$}\\
            \noalign{\smallskip}
            \hline 
            \noalign{\smallskip}
            $F=2^+ \leftarrow 2^-$ & 2494.68092 & 0.0136 \\ 
            $F=3^+ \leftarrow 2^-$ & 2494.69507 & 0.1356 \\
            $F=2^+ \leftarrow 1^-$ & 2494.73421 & 0.1221 \\
            \noalign{\smallskip}
            \hline
         \end{tabular}
\flushleft{$^{\mathrm{(a)}}$ Varberg \& Evenson (1993,
$\sigma_{\rm rms}=30$\,kHz). The frequencies for $^{18}$OH were derived by
isotope scaling. $^{\mathrm{(b)}}$ Pickett et al. (1998).}
\label{tab:lines}
\end{table}
Typical DSB receiver temperatures were 4500~K and 2500~K for the
M and L2 bands, respectively. Total power subtraction was performed by chopping 
with an amplitude of $90''$ at 1~Hz. The raw data were converted from
XFFT spectrometer (Klein et al. 2012) count rates to forward-beam
brightness temperatures with the module {\it kalibrate} (Guan et al. 2012)
as part of the {\it kosma\_software} observing software, analysing the data from
the calibration loads and the atmospheric total power and allowing us to fit
both the wet component (typically pwv=10-20~$\mu$m) and dry content of the
atmospheric emission and thus to determine the opacity correction (a few
$\times 0.1$). Further processing of the data (conversion to main-beam
brightness temperature, with a beam efficiency of 0.58, and averaging with
$1/\sigma_{\rm rms}^2$ weighting) was made with the {\sc class}
software. The overall calibration uncertainty does not exceed 20\%.
   \begin{figure}
   \centering
   \includegraphics[angle=-90,width=7.6cm]{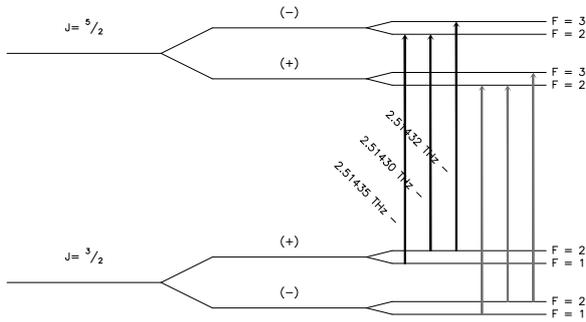}
      \caption{Level diagram (not to scale) for the $^2\Pi_{3/2}$ OH ground
               and first excited state. The observed $^{16}$OH transitions are
               indicated by bold arrows and labelled with their corresponding
               frequency. The observed $^{18}$OH transitions are drawn as grey
               arrows.}
      \label{fig:levelDiagram}
~\vspace{-0.5cm}
   \end{figure}
%
   \begin{table}
      \caption[]{Results of the OH line profile fitting.}
         \begin{tabular}{lcccc}
            \hline\hline
            \noalign{\smallskip}
            Source & $\upsilon_0^{\rm (a)}$ & {\sc fwhm}$^{\rm (b)}$ & 
                     $\tau_{\rm max}^{\rm (c)}$ & N$_{\rm OH}^{\rm (d)}$ \\ 
            \noalign{\smallskip}
 $(T_{\rm mb,c})$      & \multicolumn{2}{c}{\hrulefill~[\kms]~\hrulefill} & & [$10^{14}$~cm$^{-2}$] \\ 
            \noalign{\smallskip}
            \hline
            \noalign{\smallskip}

    W49N        & $2.9  \pm  1.9 $ & $13.4 \pm 0.4 $ & 0.9 & 
                   \multirow{2}{*}{ $8.8 \pm 4.6$ } \\
    (12.3~K)
                & $12.3 \pm  1.0 $ &$10.1 \pm 1.1 $ & 10.1 &                 \\
                & $36.8 \pm  3.5 $ &$ 11.8\pm 4.4 $ & 1.8 & $ 1.6 \pm 0.9 $\\
                & $39.1 \pm  1.3 $ &$ 2.8 \pm 0.5 $ & 2.3 & $ 0.5 \pm 0.1 $ \\
                & $60.5 \pm  0.3 $ &$10.4 \pm 3.9 $ & 2.5 & $ 2.0 \pm 0.2 $ \\
    W51         & $7.2 \pm 0.4 $ &$ 4.6\pm 2.2 $ & 2.9 & $ 0.66 \pm 0.08$ \\
    (8~K)       & $59.4 \pm 3.3  $ &$16.8\pm 4.5 $ & 5.2 & 
                         \multirow{2}{*}{$5.6 \pm 2.1 $} \\
                & $68.2 \pm 1.4  $ & $ 4.6 \pm 2.7 $ &  1.9 &  \\
    G34.26+0.15 & $12.6 \pm 0.2 $ & $ 5.7 \pm 0.8 $  & 2.5 & $ 1.12 \pm 0.21$\\
    (9~K)       & $28.3 \pm 0.5 $ & $ 6.0 \pm 1.6 $  & 1.4 & $ 0.67 \pm 0.09$\\
                & $55.4 \pm 0.4 $ & $12.9 \pm 1.0 $  & 4.8 & $ 4.73 \pm 1.20$\\
         \noalign{\smallskip}
         \hline
      \end{tabular}
\begin{list}{}{}
\item[$^{\mathrm{(a)}}$] {\sc lsr} velocity of component. The first and third
hyperfine component (hfc) are $+2.2$ and $-4.4$\,\kms, respectively, off the
second one.
\item[$^{\mathrm{(b)}}$] {\sc fwhm} of Gaussian absorption profile, deconvolved
from hfc split.
\item[$^{\mathrm{(c)}}$] Peak opacity in the strongest hfc. The saturated
absorption towards W49N (\vlsr $< 25$~\kms) provides only a lower opacity limit.
\item[$^{\mathrm{(d)}}$] Column density per fitted velocity component.
\end{list}
\label{tab:results}
~\vspace{-1.2cm}
\end{table}
%

Thanks to the high critical density of the $J=5/2 \rightarrow 3/2$ transition
($5.1\times 10^9\,{\rm cm}^{-3}$ for a 15~K gas, collision coefficients
from Dewangan et al. 1987, Einstein coefficients as given in
Tab.~\ref{tab:lines}), we can safely expect almost all OH to be in its ground 
state at the density of the foreground diffuse clouds along the sight-line.
This makes the determination of column densities much more reliable
than those derived from the $\lambda$\,18~cm line. Likewise, the
$^2\Pi_{1/2}\,J = 1/2$ level (64~K above the $^2\Pi_{3/2}\,J=5/2$ level) cannot
be substantially populated either, because we failed to detect absorption in
the simultaneously observed $^2\Pi_{1/2}\,J = 3/2 \leftarrow 1/2$ transition.
The spectral profile, in absence of emission, is thus is given by
\begin{equation}
T_{\rm mb}(\upsilon) = T_{\rm mb, c} 
\exp{\left ( -\sum_{i=1}^{N_{\rm vc}}\sum_{j=1}^{N_{\rm hfc}} \tau_\mathrm{ ij, \upsilon} \right) }\,,
\label{eq:radtran}
\end{equation}
where $T_{\rm mb}$ and $T_{\rm mb, c}$ are the main beam brightness
Rayleigh-Jeans temperatures of the spectral profile (here as a function of 
velocity) and of the continuum (in single-sideband calibration), respectively,
and $N_{\rm vc}$ and $N_{\rm hfc}$ are the number of velocity components and
hyperfine components, respectively. While uncertainties in the calibration
temperature cancel out in the opacity determination, any residual offset of 
variance $\sigma^2_{\rm rms, T{\rm c}}$ in the definition of the continuum level
leads to an additional uncertainty in the derived opacity of 
$\sigma_{\rm rms, \tau}=\sigma_{\rm rms, T_c}/T_{\rm c}$. The absorption
spectra suggest $\sigma_{\rm rms, \tau} \sim$0.1, which is tolerable in view
of the substantial opacities. A simultaneous least-squares fit to the line 
profiles of all velocity components (Eq.~\ref{eq:radtran}) with the opacity
\begin{equation}
\tau_\mathrm{ij, \upsilon} =
\sqrt{\frac{\ln{2}}{\pi}} \frac{A_{\mathrm E,j}c^3}{4\pi\Delta \upsilon_{\rm i}
\nu_{\rm j}^3}
\frac{g_{\rm u,j}}{g_{\rm l,j}} N_{\OH} w_\Lambda
\exp{
     \left(-4\ln{2}\left(\frac{\upsilon-\upsilon_{0,ij}}
                              {\Delta \upsilon_i}\right)^2\right)
    }
\end{equation}
yields $N_{\OH}$, the OH column density per velocity component. Here
$\Delta \upsilon_{\rm i}$ is the {\sc fwhm} of the Gaussian component
$i$, $g_{\rm u,j}$ and $g_{\rm l,j}$ are the statistical weights
($2F+1$) of the upper and lower level, respectively, of a given hyperfine
component $j$, $\upsilon_{\rm 0, ij}$ is the offset of its velocity from the
line-of-sight velocity of the source, and $w_\Lambda = 0.5$ corrects for the
fact that only one doublet line was observed (Fig.~\ref{fig:levelDiagram}).
Owing to the relatively large number of free parameters ($3N_{\rm vc}$), a
simulated annealing method (Metropolis algorithm) was used in combination with
a downhill simplex method (Press et al. 1992). The former assists the
minimisation process in escaping from a local minimum, and the latter improves
the efficiency of the convergence. The velocity structure in the
para-H$_2$O $1_{11} - 0_{00}$ spectrum towards W49N (Sonnentrucker et al. 2010)
suggests five velocity components as a strict minimum, while with more
components, the procedure would start to fit noise features. The results are
summarised in Tab.~\ref{tab:results}. The OH column density can be expressed by
the relationship 
$N_\OH\,\,[\mathrm{cm}^{-2}] = 7.8\times 10^{12} \tau_{\rm max} \Delta \upsilon_{\rm fwhm}$\,
[\kms] as a function of the opacity in the strongest hyperfine component and of
the width of the absorption profile of the deconvolved spectrum. The uniqueness 
of the solution was tested with a Monte Carlo study, yielding the standard
deviation of each parameter.
\section{Results}
\begin{figure}[h!]
   \centering
   \includegraphics[angle=-90,width=7.6cm]{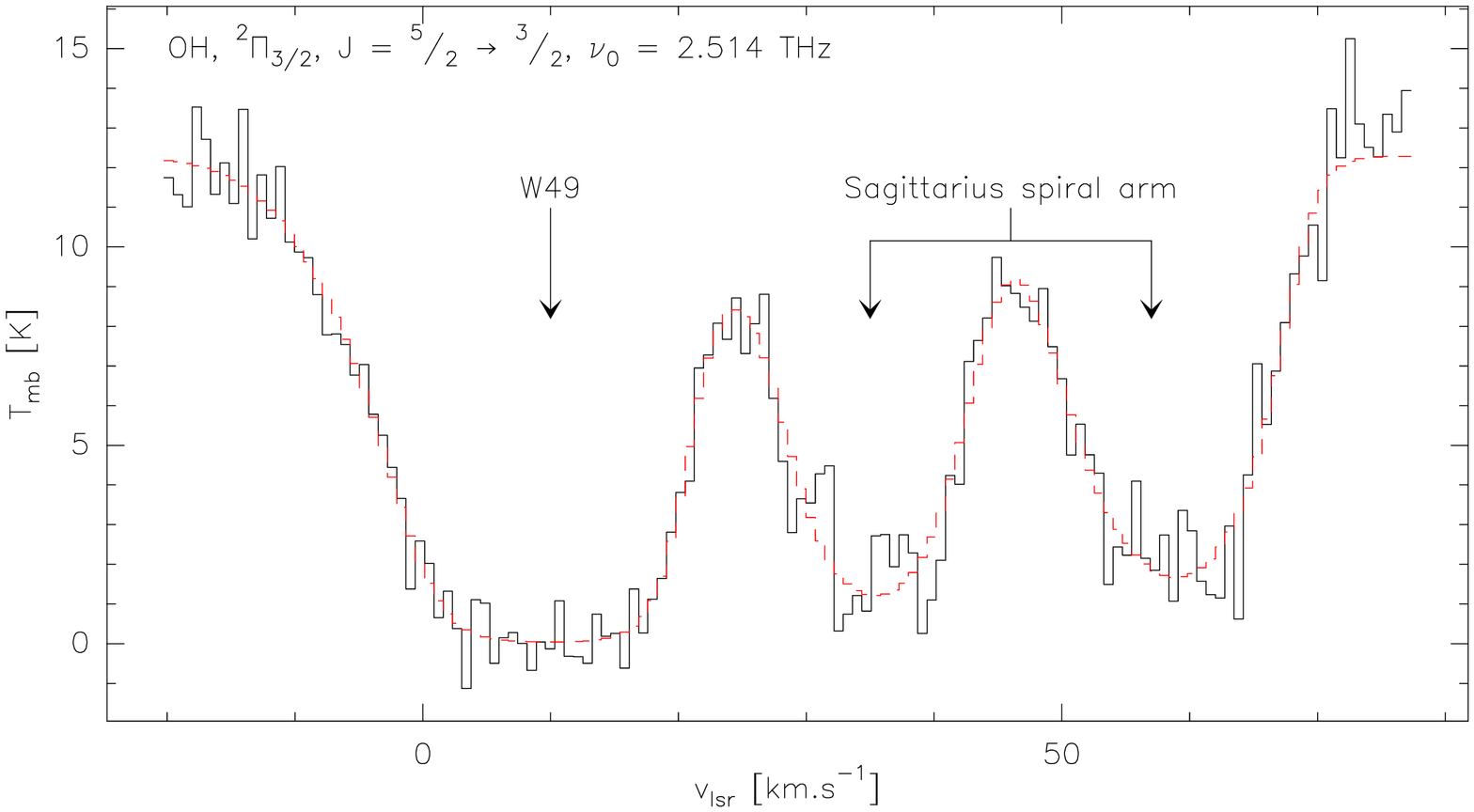}
   \includegraphics[angle=-90,width=7.6cm]{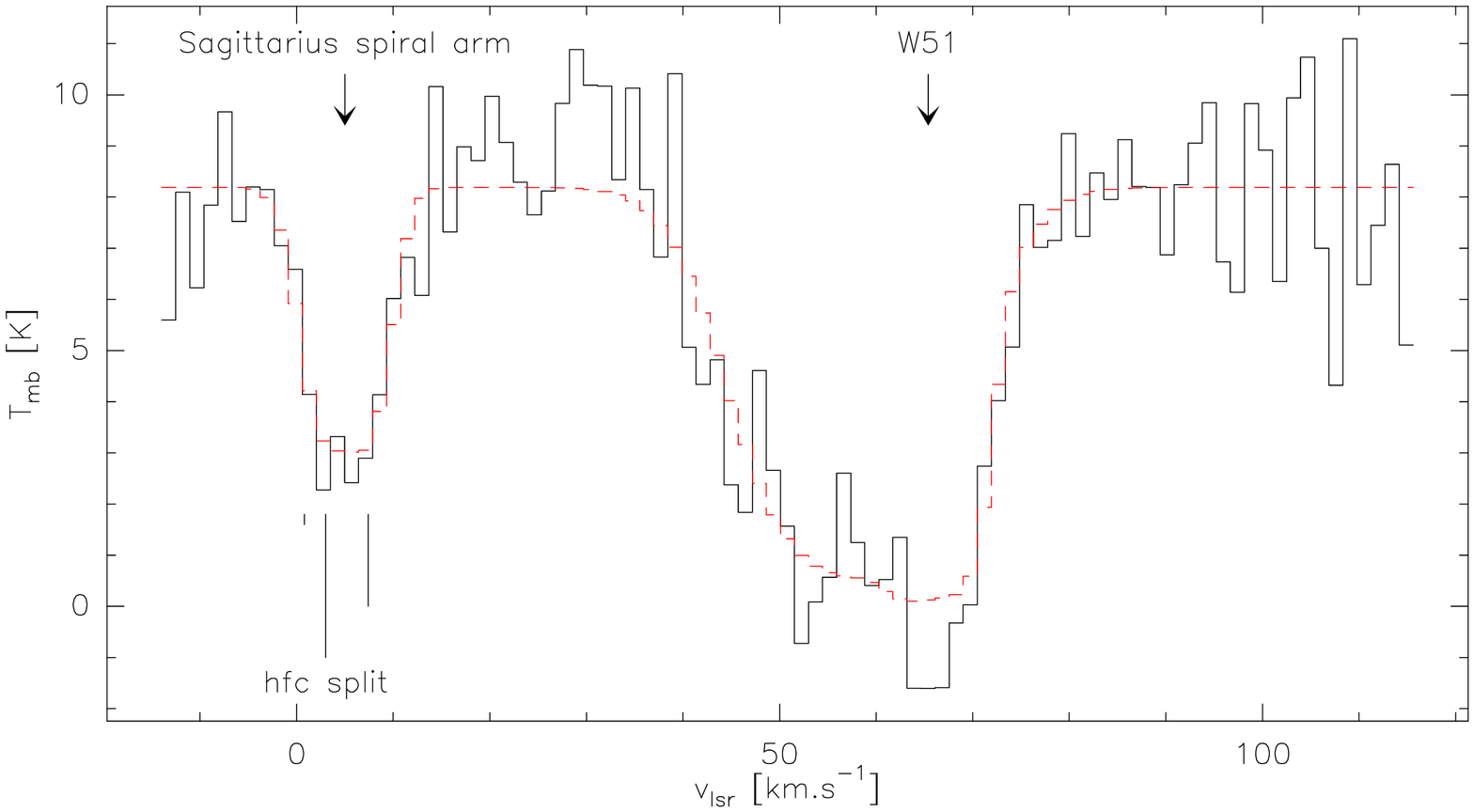}
   \includegraphics[angle=-90,width=7.6cm]{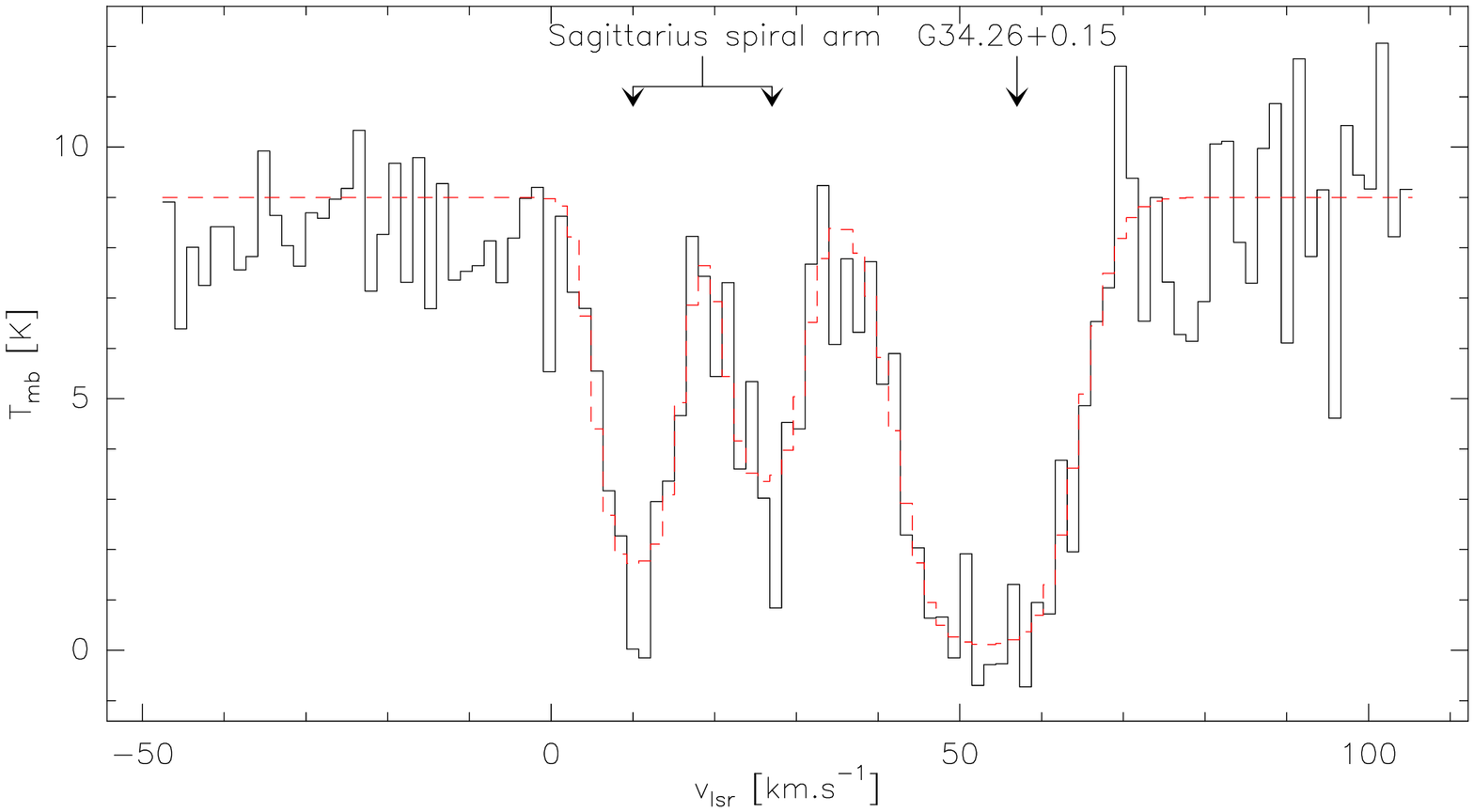}
      \caption{{\bf Top:} OH absorption towards W49N. The velocities of the
               line-of-sight clouds and of W49N are indicated. The red dashed
               line is a least-squares fit. {\bf Middle:} Same for the OH
               absorption against W51e4. The relative positions and strengths of
               the hfc splitting are indicated for the \vlsr=7~\kms~component.
               {\bf Bottom:} OH absorption against G34.26+0.15.}
   \label{fig:ohspectra}
   \end{figure}

\begin{figure}[h!]
   \centering
   \includegraphics[angle=-90,width=7.6cm]{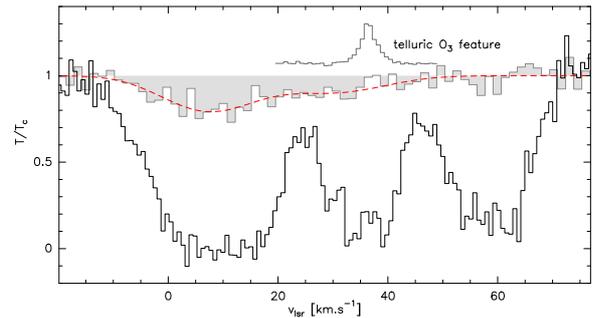}
      \caption{$^{18}$OH absorption (top, grey-shaded), a least-squares,
               two-component fit to it (dashed line) and OH absorption towards
               W49N. The spectra are scaled by the corresponding continuum
               level, to facilitate a comparison. The insert at the top shows a
               telluric ozone feature (as observed in total power), where the
               calibration is more uncertain.}
      \label{fig:w49n_18oh}
   \end{figure}

For the spiral arm clouds there is no ambiguity in the fit results 
(Fig.~\ref{fig:ohspectra}). Towards the three continuum sources the absorption
is saturated, i.e., in the $(5,20)$, $(40,80)$ and $(35,70)$~\kms~velocity
intervals for W49N, W51 and G34.26+015, respectively, where the derived main
line opacities and column densities are to be considered lower limits. For W49N,
this caveat is corroborated by the $^{18}\OH$ absorption profile
(Fig.~\ref{fig:w49n_18oh}). The $^{18}\OH/^{16}\OH$ abundance ratio is not
expected to be affected by chemical fractionation (Langer et al. 1984). The
synthesised opacity in the W49N spectrum peaks at $\tau$$=5.7$. Assuming for
the $^{18}\mathrm O/^{16}\mathrm O$ ratio the value in the 4~kpc
ring ($327 \pm 32$, Wilson \& Rood 1994; Polehampton et al. 2005 found no
evidence of an abundance gradient with galactocentric distance), the estimated
opacity in our $^{18}\OH$ detection ($\tau$$\sim$0.2) would require the main
line opacity to be higher by at least an order of magnitude with respect to
that estimated by the absorption profile fit. Unfortunately, in the
30-40\,\kms~velocity range the $^{18}\OH$ absorption is affected by a telluric
ozone feature, and a confirmation by observations of sources with a more
favourable velocity is planned to definitely rule out a baseline ripple.
In the unsaturated wings of the absorption profile, the
sensitivity of the corresponding $^{18}$OH measurement is no longer sufficient
to estimate the $^{18}\OH/\OH$ abundance ratio. A two-component fit
to the $^{18}$OH absorption (Fig.~\ref{fig:w49n_18oh}) yields a column density
of $4\times 10^{13}$\,cm$^{-2}$ for the whole absorption feature.
\begin{table*}[ht!]
\caption{OH and $\HHO$ abundances and their ratios in line-of-sight clouds
towards W49N and W51. Statistical errors are given in brackets.} 
\centering
\begin{tabular}{llcccccccc}
\hline\hline
\noalign{\smallskip}
 & & \multicolumn{5}{c}{\hrulefill~W49N \vlsr~intervals [\kms]}~\hrulefill & 
     \multicolumn{3}{c}{W51 \vlsr~intervals [\kms]} \\
\noalign{\smallskip}
\hline
\noalign{\smallskip}
    &    & (30,37) & (37,44) & (44,49) & (49,54) & (54,72) & ($-1$,11) & (11,16)$^{(c)}$&(43,50)\\
\noalign{\smallskip}
\hline
\noalign{\smallskip}
$N_\OH$ & $[10^{13}\,\mathrm{cm}^{-2}]$  & 
7.1(0.5) &11.3(6.3) & 1.2(0.2) & 1.5(0.3) & 18.1(1.3) & 6.4(0.9) & $-$ & 3.1(0.6) \\
$N_\HHO^{\rm (a)}$ & $[10^{13}\,\mathrm{cm}^{-2}]$ & 
2.3(0.1) & 6.2(0.6) & 0.1(0.05) & 1.5(0.07) & 11.6(0.4) & 2.6(0.1) & 0.23(0.05)&2.5(0.1) \\
$N_\HH^{\rm (b)}$ &$[10^{20}\,\mathrm{cm}^{-2}]$   &
5.9(1.9) & 6.6(2.0) & 0.9(0.03)&3.2(0.1) &22.6(3.7) & 4.3(0.1) & 0.6(0.03) &
2.6(0.8) \\
$[\OH]/[\HH]$& $[10^{-8}]$ & 12.0(4.0) & 17(11) & 13.2(2.5) & 4.5(0.9) & 8.0(1.4) & 15(2.1)& $-$ & 11.9(4.3)     \\
$[\HHO]/[\HH]$ & $[10^{-8}]$ & 3.9(1.3) & 9.4(3.0) & 1.1(0.6) & 4.7(0.3) & 5.1(0.9) &6.0(0.3)& 3.8(0.9) & 9.6(3.0) \\
$[\HHO]/[\OH]$ & & 0.32(0.03)& 0.55(0.31) & 0.08(0.04) & 1.03(0.20) & 0.64(0.05) & 
0.40(0.06) & $-$ & 0.81(0.16)  \\
\noalign{\smallskip}
\hline
\end{tabular}
\flushleft $^{\rm (a)}$ Sonnentrucker (2010), $^{\rm (b)}$ Godard et al. (2012). $^{\rm (c)}$ The sensitivity of the OH observations is not sufficient for this
velocity interval.
\label{tab:oh_h2o}
\end{table*}
Although the spiral arm clouds along the sight-line exhibit substantial
opacities on the order of unity, the absorption is not saturated and OH column
densities can be derived whose accuracy only depends on the signal-to-noise
ratio, the quality of the fitted profile,
and the assumed continuum level. For a comparison of our OH column densities
with those observed for $\HHO$ (Sonnentrucker et al. 2010) and inferred for
H$_2$ (Godard et al. 2012), the absorption profiles of the clouds towards W49N
and W51 are integrated within the velocity intervals of Tab.~\ref{tab:oh_h2o}.
With the exception of a velocity interval with an abnormally low water
abundance, probably due to a spectral baseline problem, the $\HHO/\OH$ ratios
are in the range 0.3-1.0. Plume et al. (2004) have determined the $\HHO/\OH$ ratio by a comparison of submillimeter
$\HHO$ observations with ground-based radio observations of the 18~cm
transitions within the ground rotational state of OH. They thereby
estimated a $\HHO/\OH$ ratio of 0.4 at \vlsr=68~\kms, in agreement with our
measurement of 0.6. The ratio of
$\sim$0.3 measured by Neufeld et al. (2002) towards W51, at \vlsr=6~\kms,
compares to our value of 0.4 in the \vlsr=($-1$,11)~\kms~interval. Their OH
column density is compatible with our value
($8\times10^{13}\,\mathrm{cm^{-2}}$ and $6.4\times10^{13}\,\mathrm{cm^{-2}}$,
respectively). Observations of the $^{2}\Pi_{1/2}\leftarrow$~$^{2}\Pi_{3/2}$
cross-band transitions towards Sgr~B2 (Polehampton et al. 2005) suggest a
$\HHO/\OH$ range of 0.6-1.2. Generally, discrepancies between different sets of
data may be due to {\sc nlte} effects in the radio lines, different spectral
resolutions and definitions of velocity components, and uncertainties in the
definition of the respective continuum levels.
%

The chemistry leading to interstellar OH and water has been 
considered in many theoretical studies over the past thirty years 
(e.g. Draine et al. 1983; van Dishoeck \& Black 1986; 
Hollenbach et al. 2009, and references therein). Three main pathways 
to OH have been identified in diffuse and translucent molecular 
clouds. The first pathway involves an ion-molecule chemistry, 
initiated by the cosmic-ray ionization of $\HH$ or H. The resulting $\HHplus$ 
and $\Hplus$ ions can lead to $\OHplus$ through the reaction sequences
$\HHplus(\HH,\H)\HHHplus(\O,\HH)\OHplus$ or
$\Hplus(\O,\H)\Oplus(\HH,\H)\OHplus$. In clouds with a low molecular fraction,
the resulting $\OHplus$ is destroyed primarily by dissociation recombination
with electrons. In clouds with a high molecular fraction, however, $\OHplus$ is
rapidly converted to $\HHHOplus$ by a series of two H atom abstraction
reactions:
$\OHplus(\HH,\H)\HHOplus(\HH,\H)\HHHOplus$. The $\HHHOplus$ ion then undergoes
dissociative recombination with electrons to form OH or $\HHO$. The branching
ratio for this process is important in determining the resultant $\OH/\HHO$
ratio and has been studied in two recent ion storage ring experiments (Jensen
et al. 2000; Neau et al. 2000): these suggest that $\sim$74\% to 83\% of
dissociative recombinations lead to OH, with almost all the remainder
leading to $\HHO$ (and less than $\sim$1\% resulting in the production of O).
In diffuse or translucent clouds, both neutral molecules are destroyed by
photodissociation, which - in the case of $\HHO$ - is an additional formation
process for OH. A second and different pathway may be important in shocks or
turbulent dissipation regions, where elevated gas temperatures can drive a
series of neutral-neutral reactions with significant energy barriers:
$\O(\HH,\H)\OH(\HH,\H)\HHO$. Finally, OH and $\HHO$ may be produced by means
of a grain-surface chemistry, in which O nuclei are hydrogenated on grain
surfaces and subsequently photodesorbed. The relative importance of these three 
pathways will determine the exact $\HHO/\OH$ abundance ratio, but all three
predict a close relationship between OH and $\HHO$. This relationship is 
supported by the observations reported here, which indicate a good 
correspondence between the OH and $\HHO$ absorption features; detailed 
modelling, which must await a larger sample of sight-lines, will be 
needed to interpret observed variations in the $\HHO/\OH$ ratio. 
We note, however, that the lower end of the observed range of $\HHO/\OH$ ratios
(0.3-1.0) is predicted by models for turbulent chemistry (Godard et al. 2009).
We note also that the observed distribution of OH is quite different from that 
of $\OHplus$; the latter is believed to arise primarily in material with a
molecular fraction that is too low to permit the efficient 
production of $\HHHOplus$, whereas the former will arise in clouds with a 
substantial abundance of $\HH$.

Future data of the OH ground state transition and the relatively high precision
of the resulting column densities will not only allow us to assess the
correlation between the abundances of OH and $\HHO$, but also to
re-calibrate less accurate OH column densities derived from decades of radio
observations.
   \begin{figure}
   \centering
   \includegraphics[angle=-90,width=7.6cm]{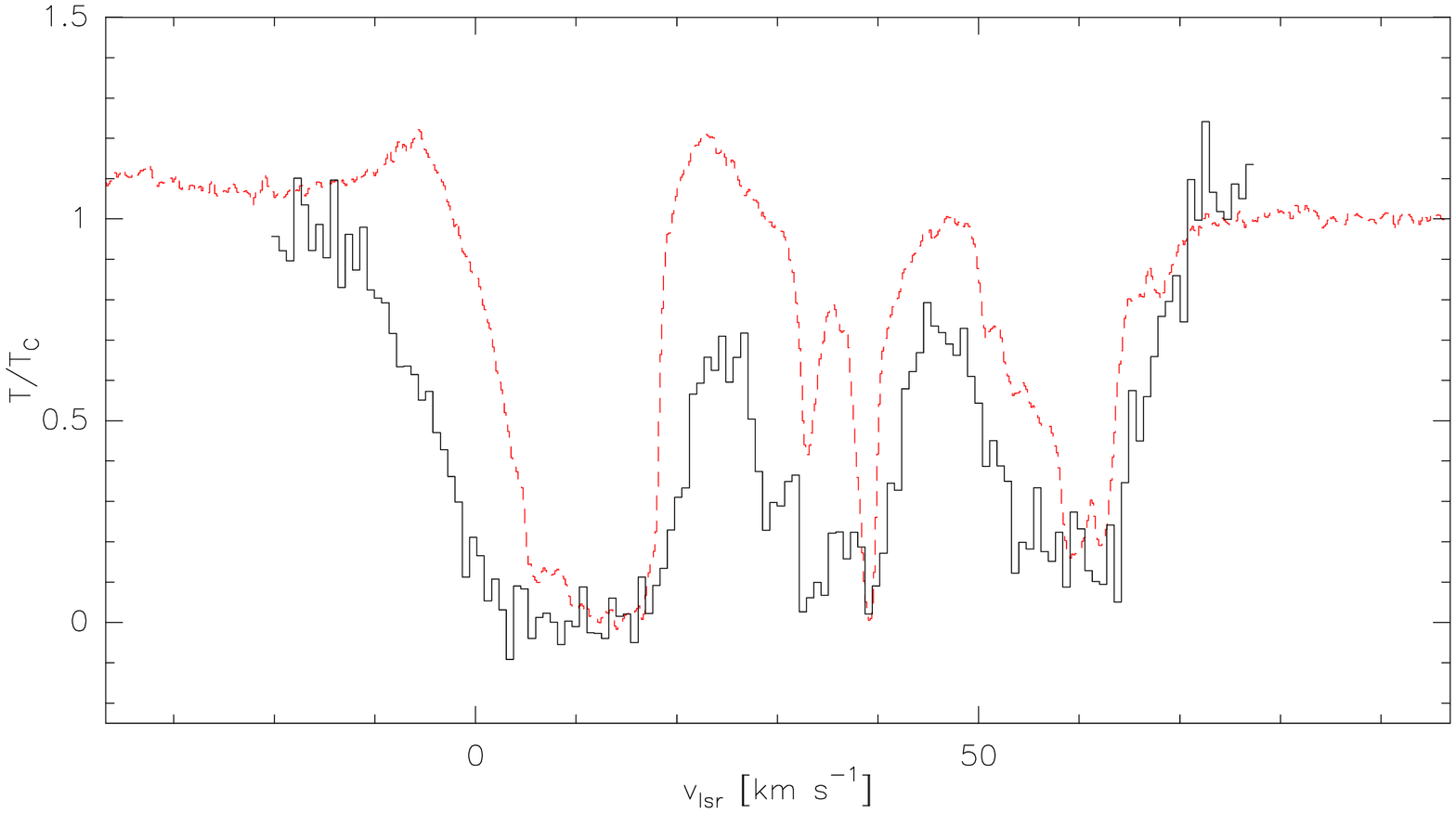}
   \includegraphics[angle=-90,width=7.6cm]{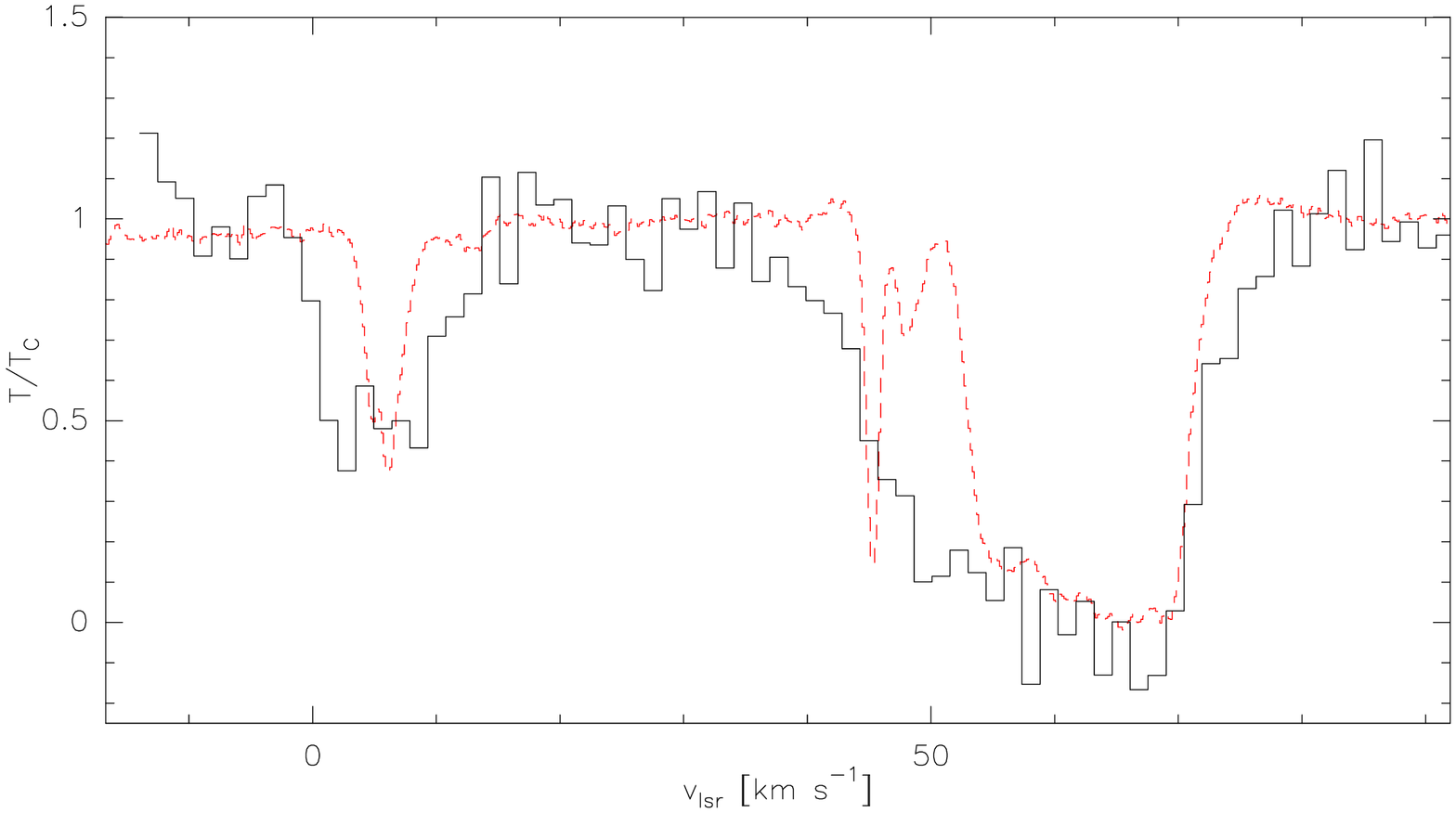}
      \caption{Para-$\HHO$ absorption (dashed red line; from Sonnentrucker et
               al. 2010) and OH absorption (solid line), normalised by the
               respective single sideband continua, towards W49N (top) and W51
               (bottom).}
      \label{fig:oh_h2o}
   \end{figure}
%
\begin{acknowledgements}
Based on observations made with the NASA/DLR Stratospheric
Observatory for Infrared Astronomy. SOFIA Science Mission Operations are
conducted jointly by the Universities Space Research Association, Inc., under
NASA contract NAS2-97001, and the Deutsches SOFIA Institut under DLR contract
50 OK 0901. We {\sc great}fully acknowledge the support by the observatory staff
and a helpful referee report.
\end{acknowledgements}

\end{document}